\RequirePackage[2021/03/19]{latexrelease}
\documentclass[aps,pra,reprint,superscriptaddress,longbibliography]{revtex4-2}

\usepackage{physics}
\usepackage{amsmath}
\usepackage{amssymb}
\usepackage{graphicx}
\usepackage{float}
\usepackage{mathrsfs}
\usepackage{mathtools}
\usepackage{extarrows}
\usepackage[margin=0.7in]{geometry}
\graphicspath{{PNG/}}
\usepackage{xcolor}

\begin{document}
	
	\title{Analytical Investigation of Focusing Bose-Einstein condensates}
	
	\author{R.~Richberg}
	
	\email{formerly known as Amir M. Kordbacheh}
	
	\affiliation{ 
		Department of Quantum Science, Research School of Physics, The Australian National University, Canberra, ACT 2601, Australia
	}
	
	\affiliation{ 
		School of Physics, University of Melbourne, Melbourne, 3010, Australia
	}%

	\author{A.~M.~Martin}%
	
	\affiliation{ 
		School of Physics, University of Melbourne, Melbourne, 3010, Australia
	}%

	\date{\today}

	\begin{abstract}
		
		The focusing of a propagating untrapped Bose-Einstein condensate is studied theoretically. We use a scaling solution method comprising a time-dependent scaling function to analytically examine the dynamics of a falling Bose-Einstein condensate in different regimes of propagation including the expansion and compression zones. Our model is based on the Gross-Pitaevskii equation which involves the interparticle interactions between atoms, and consequently their influence on the focused structures. We investigate the focused profile characteristic factors such as the resolution and peak density for various cases of the focusing optical potential parameters as well as the factors associated with the moving cloud. Our results are compared with numerical solutions of the Gross-Pitaevskii equation.

	\end{abstract}
	
	\maketitle

	\section{Introduction}
	
The construction of atom scale devices using multiple electronic materials in a single step printing process with high throughput represents the ultimate fabrication capability. Such implementation enables the design and test of quantum three-dimensional (3D) computer chips and processors \cite{3, 54}. The focusing of neutral atomic beams using optical laser lights in order to create nano-structure sizes has been established in \cite{I43, I42, I41, I44, 2_17, I45, I46, I47, I48, I49, I50, I51}. Taking advantage of optical lattices (standing waves) results in a large array of identical substructures with a high spatial coherence such as photonic materials \cite{RS_1}. As a case in point, efficient sensors are produced by an array of uniform nanostructures covering a significant area \cite{RS_1}. In addition, focusing a beam of neutral atoms in atom optics can create nanoscale metal dots on a surface \cite{RS_2}, which are utilized to study transport phenomena, or quantum dot effects when deposited on a semiconductor. Another significant application is the use of metal dots as an etch mask \cite{RS_3} to transfer the pattern to a substrate material, allowing the extension of the fabrication techniques to other materials.

A vast majority of research conducted in the scope of atom lithography to date has exploited oven sources of neutral atoms, which restricts structures to a range of $60$-$100$ nm. However, using controllable matter-waves such as Bose-Einstein condensates (BECs) for atom lithography \cite{I41, 2_17} offers several advantages over thermal atom sources, including smaller de Broglie wavelengths, higher peak densities, higher quality spatial modes, and superior coherence \cite{3_2, 3_3, I40}. A comparison between oven and BEC sources indicates that while thermal sources may produce a flux range that is orders of magnitude higher than condensate sources, they have limited structural resolutions due to angular divergence. Nevertheless, an ultra cold source of atoms such as a BEC enables spatial modes leading to collimated atomic beam along with an enhanced flux density \cite{3_3}. 	
  
Regarding the fact that the neutral alkali atoms such as Li, Na, K, Rb and Cs are chemically highly active elements and could be simply manipulated by the laser fields, they are an appropriate choice for atom lithography \cite{I75}. In our study, we investigate the properties and physics of atom deposition technique utilizing a BEC source of $^{87}$Rb since it is a well-studied system along with defined and measured experimental parameters for the purpose of analytical and numerical computations.

The examination of focusing of a confined BEC in a harmonic trap was conducted theoretically in \cite{I6} where the focusing time was scaled as a function of different focusing strengths. Later in 2010, Judd {\it et al}. \cite{I74} studied the evolution of a compressed BEC utilizing Frensel zone plates (FZPs), which predicted a resolution of $50$ nm. Recently, using an optical lattice and a harmonic focusing potential, the focusing dynamics of freely propagating BECs was considered via Gross-Pitaevskii equation (GPE) simulations \cite{RS_4, RS_5} as well as the analytical variational method \cite{RS_6, RS_7} in which the profile linewidths were predicted to be as narrow as $20$ nm in the former, and $9$ nm in the latter. In a similar work, the focusing of a quasi-continuous atom laser beam of $^{85}$Rb was studied; employing a two-state model which involves the two-body atom-atom interactions and three-body recombination losses, the resolution of $8$ nm was predicted \cite{RS_8}. 
 
 In this paper, we will introduce a scaling solution method \cite{4_1} to consider the dynamics of an untrapped propagating BEC under an externally optical focusing potential. Applying this analytical approach, we aim to understand the effect of inter-atomic interactions within the BECs on the deposited structures. To this end, factors such as the intensity and geometry of the focusing light field as well as various magnitudes of imparted momentum kicks to the BECs in the atom deposition process are investigated. The quality of the focussing is assessed by examining the width of the BEC and their resultant atomic density at the focus. Ultimately, comparing the analytical results with the corresponding numerical GPE simulations, we evaluate the validity of prediction in our model for a focused BEC.

	\section{The Scaling Solution Model}
\label{sec:S1}

In \cite{RS_5}, the GPE was used to describe the focusing dynamics of a repulsively-interacting BEC in a Gaussian standing potential. However, in general the GPE is a 3D partial differential equation that requires a numerical solution, and large grid sizes (particularly for free-space dynamics). This restricts the development cycle, making rapid prototyping of focusing protocols unachievable. Hence, we consider a scaling solution approach \cite{4_1} to estimated the wavefunction of the BEC in different regimes during its propagation. We adapt this methodology to account for an evolving falling BEC, released from a trap, being eventually focused by a harmonic focusing potential.

Let us consider the condensate wavefunction as $\psi_0(x, y, z, t)$. It is assumed that the condensate is initially confined at $t=0$ by a trapping potential with a cylindrical symmetry including two radial tight frequencies along the $y$ and $z$ axes, $\omega_{0y}=\omega_{0z}=\omega_{0r}$, and one axial weak frequency along the $x$ axis, $\omega_{0x}$ where $\omega_{0x}<\omega_{0r}$. This causes a cigar-shaped condensate elongated along the $x$ axis. Once the trap is turned off at $t>0$, the BEC is exposed to an external time-dependent harmonic potential, $V(x, r, t)$. Therefore, the whole potential function affecting the BEC is represented by 
\begin{equation}
	V(x,r,t)=
		\frac{1}{2}m(\omega_{x}^2(t)x^2+\omega_{r}^2(t)r^2)
			\label{428}
\end{equation}
where $\omega_x(t=0)=\omega_{0x}$, $\omega_r(t=0)=\omega_{0r}$, $r^2=y^2+z^2$, $\omega_{x}(t)$ and $\omega_{r}(t)$ are, respectively, the time-dependent axial and radial frequencies of the optical harmonic focusing potential. The evolution of the condensate wavefunction, $\psi_0(x, r, t)$, in such a potential is described by the Gross-Pitaevskii equation
\begin{equation}
	\begin{split}
	i\hbar\frac{\partial\psi_0(x,r,t)}{\partial t}=\Big(-\frac{\hbar^2}{2m}(\nabla_x^2+\nabla_r^2)+\frac{m}{2}\big(\omega_x^2(t)x^2+\omega_r^2(t)r^2\big)\\
	+g|\psi_0(x, r, t)|^2\Big)\psi_0(x,r,t),~~~~~~~~~~~~~~~~~~~
	\label{429}
	\end{split}
\end{equation}
where $m$ and $\hbar$ are the atomic mass and Planck's constant, $g=\frac{4\pi\hbar^2a_s}{m}$ measures the interaction strength, $a_s$ is the scattering length, $\nabla_x^2=\partial^2/\partial x^2$ and $\nabla_r^2=\partial^2/\partial y^2+\partial^2/\partial z^2$.

Since the harmonic frequencies vary over time, the time and distance scales are changed such that the rescaled coordinates are turned to $\rho_x=x/b_x(t)$, $\rho_r=r/b_r(t)$ where $b_x(t)$ and $b_r(t)$ are the dimensionless scaling factors, and the rescaled time is introduced by $\tau(t)$ \cite{4_1,4_3}. In this case, the wavefunction of the cigar-shaped condensate is given by: 
\begin{equation}
	\psi_0(x, r, t)=\frac{1}{\sqrt{K(t)}}\chi_0(\rho_x, \rho_r, \tau(t))\exp\{i\Phi(r, x, t)\},
	\label{430}
\end{equation}
where $K(t)=b_x(t)b_r^2(t)$ is a dimensionless quantity, and the dynamical phase is defined by
\begin{equation}
	\Phi(x, r, t)=\frac{m}{2\hbar}\Bigg(\frac{\dot{b}_x(t)}{b_x(t)}x^2+\frac{\dot{b}_r(t)}{b_r(t)}r^2\Bigg).
	\label{431}
\end{equation}
The total number of atoms in the rescaled system within the BEC is achieved by the normalization condition $\int|\chi_0|^2d\rho_xd^2\rho_r=\int|\psi_0|^2dxd^2r=N_0$. The GPE for the rescaled wavefunction $\chi_0(\rho_x, \rho_r, \tau(t))$, is gained by substituting Eqs. (\ref{430}) and (\ref{431}) into Eq. (\ref{429}),
\begin{widetext}
\begin{equation}
	i\hbar\frac{d\tau}{dt}\frac{\partial\chi_0}{\partial\tau}=\Bigg[-\frac{\hbar^2}{2m}\Big(\frac{1}{b_x^2(t)}\nabla^2_{\rho_x}+\frac{1}{b_r^2(t)}\nabla^2_{\rho_r}\Big)
	+\frac{m\rho^2_x}{2}\Big(\ddot{b}_x(t)b_x(t)+\omega_x^2(t)b^2_x(t)\Big)+\frac{m\rho^2_r}{2}\Big(\ddot{b}_r(t)b_r(t)+\omega_r^2(t)b^2_r(t)\Big)+\frac{g|\chi_0|^2}{K(t)}\Bigg]\chi_0.
	\label{405}
\end{equation}
\end{widetext}
We now break down $\chi_0(\rho_x, \rho_r, \tau(t))$, into the time and position scales,
\begin{equation}
	\chi_0(\rho_x, \rho_r, \tau(t))=\eta(\rho_x, \rho_r)\exp\big(-i\mu\tau(t)/\hbar\big),
	\label{406}
\end{equation}
where $\mu$ is the chemical potential, defined by \cite{3_5}
\begin{equation}
	\mu=\frac{1}{2}\hbar\overline{\omega}_0\Big(15N_0a_s\sqrt{\frac{m\overline{\omega}_0}{\hbar}}\Big)^{2/5},
	\label{1919}
\end{equation}
where $\overline{\omega}_0=[\omega_{0x}\omega_{0y}\omega_{0z}]^{1/3}$. Using Eq. (\ref{406}) as well as applying the Thomas-Fermi (TF) approximation \cite{3_5}, Eq. (\ref{405}) reduces to
\begin{equation}
	\begin{split}
	\frac{m\rho^2_x}{2}\Big(\ddot{b}_x(t)b_x(t)+\omega_x^2(t)b^2_x(t)\Big)~~~~~~~~~~~~~~~~~~~~~~~~~~~~~~~~~~~\\
	+\frac{m\rho^2_r}{2}\Big(\ddot{b}_r(t)b_r(t)+\omega_r^2(t)b^2_r(t)\Big)+\frac{g|\eta|^2}{K(t)}-\mu\frac{d\tau}{dt}=0.
	\label{408}
	\end{split}
\end{equation}
The scaling equations for $b_x(t)$, $b_r(t)$ and $\tau(t)$ are then obtained by the following choices,
\begin{equation}
	\ddot{b}_x(t)+\omega_x^2(t)b_x(t)=\frac{\omega_{0x}^2}{b_x(t)K(t)};
	\label{432}
\end{equation}
\begin{equation}
	\ddot{b}_r(t)+\omega_r^2(t)b_r(t)=\frac{\omega_{0r}^2}{b_r(t)K(t)};
	\label{433}
\end{equation}
\begin{equation}
	\tau(t)=\int^t\frac{dt'}{K(t')},
	\label{436}
\end{equation}
where the initial conditions for the scaling factors are given by
\begin{equation}
	b_x(0)=b_r(0)=1;
	\label{434}
\end{equation}
\begin{equation}
	\dot{b}_x(0)=\dot{b}_r(0)=0.
	\label{435}
\end{equation}
Substituting Eqs. (\ref{432})~-~(\ref{436}) into Eq. (\ref{408}), this reduces to
\begin{equation}
	\frac{m}{2}\Big(\rho^2_x\omega^2_{0x}+\rho^2_r\omega^2_{0r}\Big)+g|\eta|^2-\mu=0.
	\label{413}
\end{equation}
Eq. (\ref{413}) results in the Thomas-Fermi wavefunction in terms of the rescaled coordinates, 
\begin{equation}
	\eta(\rho_x, \rho_r)=\sqrt{\frac{\mu}{g}\bigg(1-\frac{\rho_x^2}{R_{x}^2}-\frac{\rho_r^2}{R_{r}^2}\bigg)}~,
	\label{439}
\end{equation}
where $R_{x}=\sqrt{\frac{2\mu}{m\omega_{0x}^2}}$ and $R_{r}=\sqrt{\frac{2\mu}{m\omega_{0r}^2}}$ are, respectively, the condensate TF radii along the axial and radial trap axes at $t=0$. By defining $R_x(t)=b_x(t)R_{x}$ and $R_r(t)=b_r(t)R_{r}$, which indicate the evolving TF radii at $t>0$, the TF wavefunction in the original cylindrical coordinates is achieved   
\begin{equation}
	\eta\bigg(\frac{x}{b_x(t)}, \frac{r}{b_r(t)}\bigg)=\sqrt{\frac{\mu}{g}\bigg(1-\frac{x^2}{R^2_x(t)}-\frac{r^2}{R^2_r(t)}\bigg)}~.
	\label{4391}
\end{equation}
Hence, the general symbolic wavefunction of the evolving BEC is obtained using Eqs. (\ref{430}), (\ref{431}), (\ref{406}) and (\ref{4391}),
\begin{equation}
	\begin{split}
	\psi_0(x, r, t)=\frac{1}{\sqrt{K(t)}}\eta\bigg(\frac{x}{b_x(t)}, \frac{r}{b_r(t)}\bigg)
	~~~~~~~\\ \times\exp\{i\Phi(x, r, t)-i\mu\tau(t)/\hbar\}.
	\label{440}
	\end{split}
\end{equation}

\begin{figure}[tb]
	\centering
	\includegraphics[width=9.5cm, height=8.5cm,angle=0]{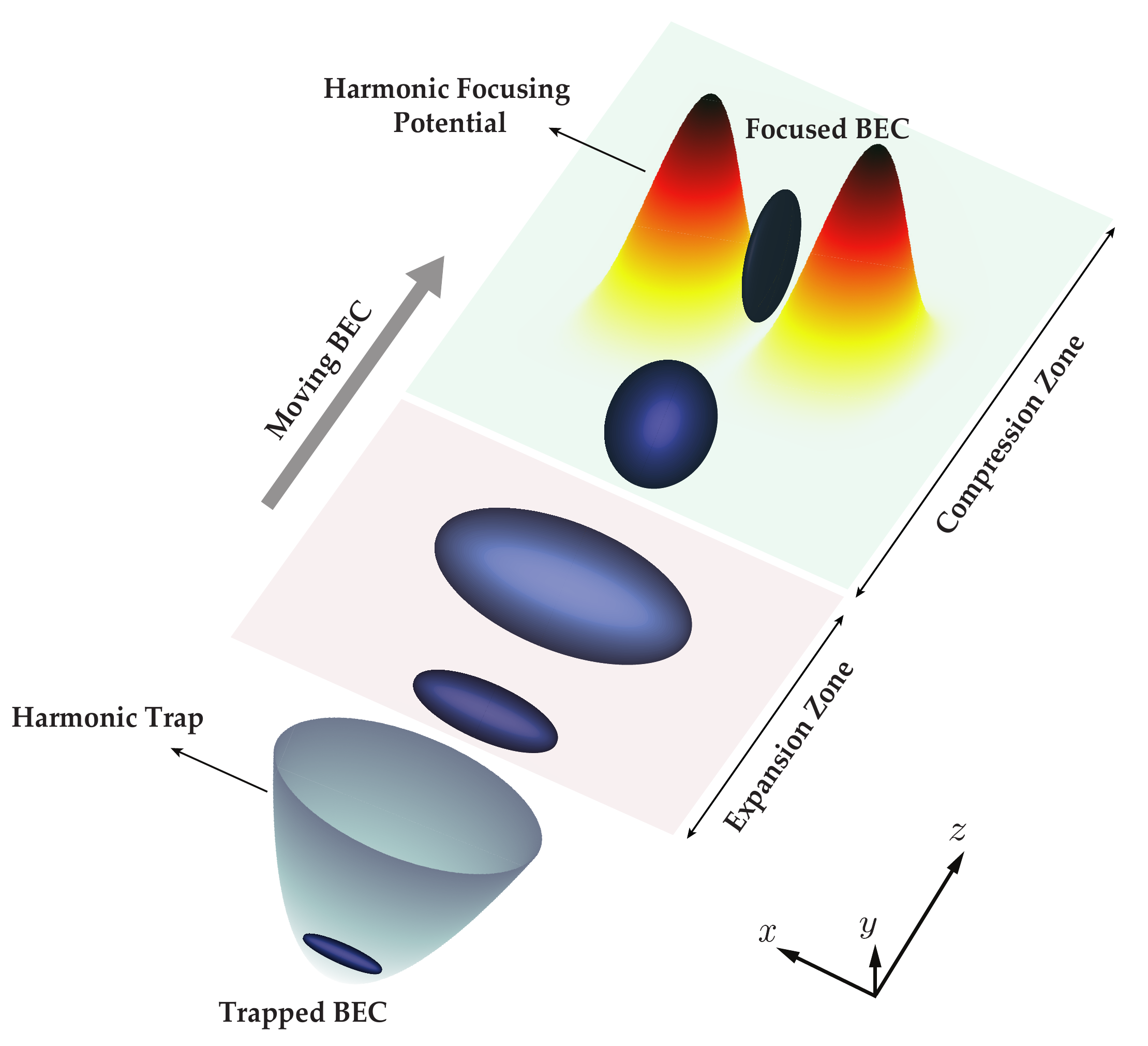}
	\caption{Schematic illustration of the focusing process through an optical harmonic potential. The BEC is released from a trapping potential, indicated by the 3D parabola, and propagates and expands along the $z$-axis in the expansion zone indicated by the light red rectangle. The compression zone is depicted by a light green rectangle in which the BEC is focused as it approaches the harmonic focusing potential, which is indicated by the two parallel yellow-red hills. A complete focused BEC occurs between the two adjacent peaks of the focusing potential, which has a harmonic distribution along the focusing $x$ axis. The problem has a symmetry along the $y$ axis.}
	\label{f1}
\end{figure}


In order to estimate the scaling functions, we split the problem into two regimes, the expansion and compression zones, as depicted in Fig~\ref{f1}. Initially, at $t=0$, the condensate is trapped by a harmonic trapping potential. At $t>0$, the condensate is released from the trap and starts expanding freely whilst moving towards the harmonic focusing potential along the $z$ axis. The BEC then enters the second regime where it undergoes the action of a focusing potential and starts being compressed until it is optimally focused by the potential. 

For the case of free expansion in the time interval $0<t\leqslant t_0$ ($t_0$ is the end time of expansion), and given $\omega_x(t)=\omega_r(t)=0$ in the expansion zone, Eqs. (\ref{432}) and  (\ref{433}) reduce to
\begin{equation}
	\ddot{b}_{xe}(t)=\frac{\omega_{0x}^2}{b_{xe}^2(t)b_{re}^2(t)};
	\label{441}
\end{equation}
\begin{equation}
	\ddot{b}_{re}(t)=\frac{\omega_{0r}^2}{b_{re}^3(t)b_{xe}(t)},
	\label{442}
\end{equation}
where the index $e$ refers to the expansion case. Given that the condensate is initially elongated along the $x$ axis [$\omega_{0r}\gg\omega_{0x}$, or $R_r(0)\ll R_x(0)$], the expansion along the radial direction would be much faster once the BEC is released from the trap. As a result, at $t<t_0$, one can consider $b_{xe}(t)$ as a constant variable over the expansion process, meaning that $b_{xe}(t)=b_{xe}(0)=1$. Applying this condition, the solution for (\ref{442}) reads
\begin{equation}
	b_{re}(t)=\sqrt{1+\omega_{0r}^2t^2},
	\label{443}
\end{equation}
which then is substituted into (\ref{441}) and leads to 
\begin{equation}
	b_{xe}(t)=1+\bigg(\frac{\omega_{0x}}{\omega_{0r}}\bigg)^2\bigg[(\omega_{0r}t)\tan^{-1}(\omega_{0r}t)-\ln\big(\sqrt{1+(\omega_{0r}t)^2}\big)\bigg].
	\label{444}
\end{equation}
It is noted that Eqs. (\ref{441}) and (\ref{442}) can also be solved numerically, which consequently results the numerical solution of the rescaled expansion time function, $\tau_e(t)$ [see Eq. (\ref{436})]. Finally,  Eq. (\ref{440}), specifically for the expanding BEC in $t_0<t\leqslant t_1$, converts to
\begin{equation}
	\begin{split}
	\psi_e(x, r, t)=\frac{1}{\sqrt{b_{xe}b_{re}^2}}\sqrt{\frac{\mu}{g}\bigg(1-\frac{x^2}{R_{xe}^2(t)}-\frac{r^2}{R_{re}^2(t)}\bigg)}~~~~~\\
	\times\exp\Bigg[\frac{im}{2\hbar}\bigg(\frac{\dot{b}_{xe}}{b_{xe}}x^2+\frac{\dot{b}_{re}}{b_{re}}r^2-\frac{2\mu\tau_{e}}{m}\bigg)\Bigg].
	\label{445}
	\end{split}
\end{equation}

Turning to the condensate compression process, the focusing potential is switched on at $t>t_0$ and it varies in the time interval $t_0<t\leqslant t_1$ ($t_1$ is the end of compression process where the BEC is optimally focused). We assume that the dipole force gradient \cite{37} caused by the focusing potential, is only applied along the $x$ axis, and there exists no other external forces exerted to the BEC along the radial, $y$ and $z$ axes such that $\omega_x(t_0)=0$, $\omega_x(t>t_0)>0$ and $\omega_r(t_0)=\omega_r(t>t_0)=0$. The choice for the focusing potential configuration will be discussed in Section~\ref{sec:S3}. As a result, Eqs. (\ref{432}) and (\ref{433}) change to
\begin{equation}
	\ddot{b}_{xc}(t)+\omega_x^2(t)b_{xc}(t)=0;
	\label{446}
\end{equation}
\begin{equation}
	\ddot{b}_{rc}(t)=0,
	\label{447}
\end{equation}
where the index $c$ refers to the compression case. The solution for $b_{rc}(t)$ is simply obtained by integrating twice from both sides of Eq. (\ref{447}),
\begin{equation}
	b_{rc}(t)=Ct+D,
	\label{448}
\end{equation}
where the constant coefficients, $C$ and $D$, are determined using the boundary conditions $b_{re}(t_0)=b_{rc}(t_0)=b_{0re}$ and $\dot{b}_{re}(t_0)=\dot{b}_{rc}(t_0)=\dot{b}_{0re}$
\begin{equation}
	C=\dot{b}_{0re};
	\label{449}
\end{equation}
\begin{equation}
	D=b_{0re}-\dot{b}_{0re}t_0.
	\label{450}
\end{equation}
Since $\omega_x(t)$ is a time-dependent function, it makes Eq. (\ref{446}) complicated to solve analytically. Hence, it is reasonable to estimate $b_{xc}(t)$ via a numerical solution of Eq. (\ref{446}) in the time interval $t_0<t\leqslant t_1$ using the above-mentioned boundary conditions. The rescaled compression time function, $\tau_c(t)$, is obtained consequently by a numerical integration of Eq. (\ref{436}). Overall, Eq.(\ref{440}), for the compression process, in $t_0<t\leqslant t_1$, is rewritten as
\begin{equation}
	\begin{split}
	\psi_c(x, r, t)=\frac{1}{\sqrt{b_{xc}b_{rc}^2}}\sqrt{\frac{\mu}{g}\bigg(1-\frac{x^2}{R_{xc}^2(t)}-\frac{r^2}{R_{rc}^2(t)}\bigg)}
	~~~~~~\\ \times\exp\Bigg[\frac{im}{2\hbar}\bigg(\frac{\dot{b}_{xc}}{b_{xc}}x^2+\frac{\dot{b}_{rc}}{b_{rc}}r^2-\frac{2\mu\tau_c}{m}\bigg)\Bigg].
	\label{451}
	\end{split}
\end{equation}

\section {The Focusing Potential}
\label{sec:S3}

In order to investigate the compression dynamics of the BEC, one needs to determine the required focusing time-dependent frequency. We assume two counter-propagating laser beams creating a focusing light field along the $x$-axis. The BEC propagates along the $z$-axis perpendicular to the propagation of the light field. Atoms are assumed to move slowly enough inside the light field to avoid the spontaneous emission, and they also maintain the adiabatic conditions \cite{2_7}. In this case, the BEC is influenced by the dipole force \cite{2_2} along the $x$ axis. The corresponding dipole potential resulting from the interaction between an induced atomic dipole moment and the light electric field, experienced by the moving BEC \cite{2_6, RS_5, RS_7, RS_8} is given by
\begin{equation}
	U_{\text{dip}} (x,z)=\frac{\hbar\Delta}{2}\ln\Bigg(1+\frac{\gamma^2}{\gamma^2+4\Delta^2} \frac{I(x,z)}{I_s}\Bigg),
	\label{201}
\end{equation}
which depends on factors such as spontaneous decay rate of the excited state, $\gamma$, the detuning from the resonance $\Delta$, and the saturation intensity, $I_s$, associated with the atomic D$_2$ line, $5~^2S_{1/2}\longrightarrow 5~^2P_{3/2}$, for $^{87}$Rb. The potential intensity profile, $I(x, z)$, is chosen to be harmonic shaped along the $x$ axis with a single node at $x=0$, which can be practically produced using a spatial light modulator \cite{38,39}. In addition, the profile comprises a Gaussian distribution along the $z$ axis resulting in,
\begin{equation}
	I(x,z)=I_0 \exp(-2z^2/\sigma_z^2)(k^2x^2),
	\label{452}
\end{equation}
where $I_0$ is the maximum intensity of the spatially varying harmonic profile, $\sigma_z$ is the radius of the beam at $1/e^2$ value of the maximum intensity, $k=2\pi/\lambda$ determines the strength of the potential and $\lambda$ is the wavelength of the field.	In such a case, the light forces applied to moving atoms along the transverse $y$-axis are negligible compared to those along the $x$-axis \cite{2_7}. As a result, a translational symmetry is formed along the $y$-axis.

For the purpose of simplicity in the computational process, the BEC is assumed to be situated in a stationary frame while the harmonic focusing potential is located in a moving frame approaching towards the BEC \cite{RS_5, RS_7}. Furthermore, considering relatively low values of $I_0$, and relatively large values of $\Delta$, Eq. (\ref{201}) converts to
\begin{equation}
	U_{\text{dip}}(x,t)=\frac{\hbar\Delta\gamma^2}{(\gamma^2+4\Delta^2)}\frac{I_0}{I_s}k^2x^2f(t).
	\label{455}
\end{equation}
where 
\begin{equation}
	f(t)=\exp\Big(\frac{-2}{\sigma_z^2}\big[z_0-z(t)\big]^2\Big),
	\label{454}
\end{equation}
where $z_0$, the initial distance between the centre-of-mass of the condensate and the center of the focusing potential, $z(t)=\frac{1}{2}gt^2+v_0t$ is the varying distance in terms of time, following the free falling motion, and $g$ denotes the gravity. Since Eq. (\ref{455}) has an harmonic configuration, one can write 
\begin{equation}
	V(x,t>0)=\frac{1}{2}m\omega_x^2(t)x^2=U_{\text{dip}}(x,t),
	\label{480}
\end{equation} 
where the time-dependent focusing frequency along the $x$ axis is achieved as
\begin{equation}
	\omega_x^2(t)=\frac{\hbar\Delta\gamma^2k^2}{m(\gamma^2+4\Delta^2)}\frac{I_0}{I_s}\exp\Bigg\{\frac{-2}{\sigma_z^2}\Bigg[z_0-\Bigg(\frac{1}{2}gt^2+v_0t\Bigg)\Bigg]^2\Bigg\}.
	\label{456}
\end{equation}

Eq. (\ref{456}) implies that the focusing potential, propagating along the $x$ axis, is configured as a time-dependent function in $0<t<\infty$ with a Gaussian distribution starting from zero to its peak intensity, then falling back to zero along the $z$ axis. Hence, it is reasonable to consolidate the compression and expansion states, and numerically solve the scaling factors, $b_x(t)$ and $b_r(t)$, directly from Eqs. (\ref{432}) and (\ref{433}) for $0<t<\infty$ considering the time-dependent frequency from Eq. (\ref{456}). Accordingly, the BEC wavefunction is scaled by Eq. (\ref{440}) at any $t$ from the solution of $b_x(t)$ and $b_r(t)$.

\begin{figure}[t!]
	\centering
	\includegraphics[width=8cm, height=6.5cm,angle=0] {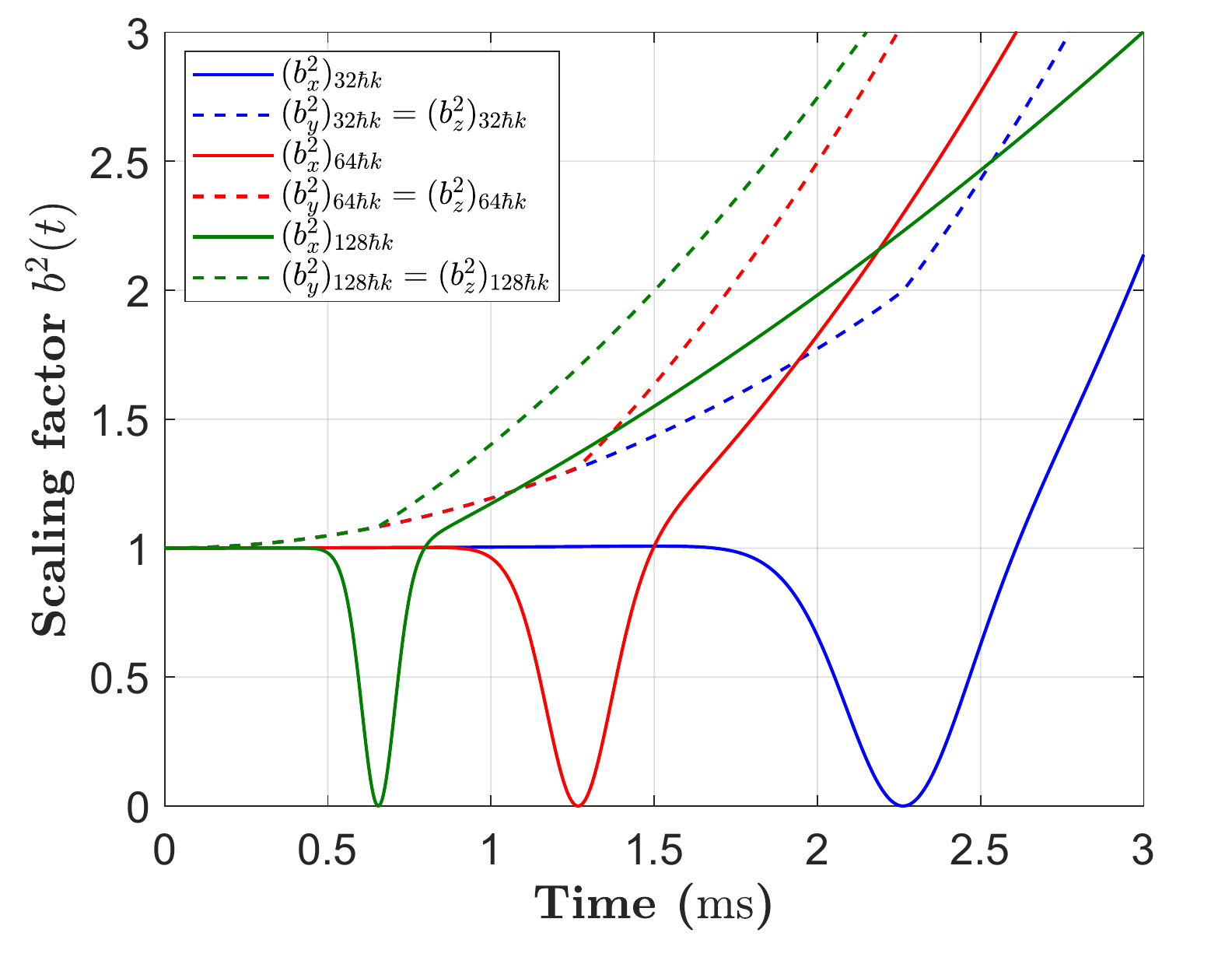}
	\caption{Numerical solution for the scaling factor (associated with the $^{87}$Rb BEC) as a function of time for three momentum kicks, $p=32 \hbar k$ (solid and dashed blue curves), $p=64 \hbar k$ (solid and dashed red curves) and $p=128 \hbar k$ (solid and dashed green curves). While the solid curves indicate the evolution of the scaling factor along the $x$ axis, $b_x^2(t)$, the dashed ones show this procedure along the radial $y$ and $z$ axes by $b_y^2(t)$ and $b_z^2(t)$. The parameters used in this simulation are: $z_0=500~\mu$m, $\sigma_z=100~\mu$m, $\lambda=312~\mu$m, $\Delta=200$ GHz, $I_s=16.5$ W/m$^2$, $\gamma=37$ MHz, $\omega_{0x}= 2\pi\times 10$ Hz and $\omega_{0y}=\omega_{0z}=2\pi\times 70$ Hz.}
	\label{f401}
\end{figure}

\begin{figure}[t!]
	\centering
	\includegraphics[width=9cm, height=12.5cm,angle=0] {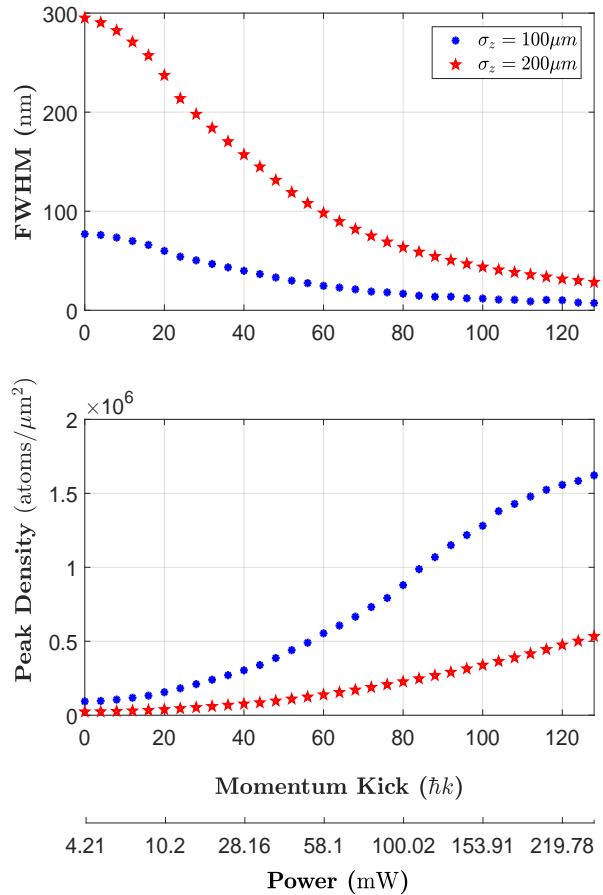}
	\caption{Results for the focused $^{87}$Rb BEC at $z=0$ derived from the scaling solution model for different values of momentum kick and focusing power while considering two different potential beam sizes, $\sigma_z=100~\mu$m (indicated by the blue dots) and $\sigma_z=200~\mu$m (indicated by the red triangles). The upper graph illustrates the values of FWHM and the lower one shows the corresponding peak densities. The horizontal axis at the bottom of the figure represents the corresponding optimal power values to the momentum kicks.
		Parameters used are: $N_0=10^5$, $z_0=500~\mu$m, $\lambda=312~\mu$m, $\Delta=200$ GHz, $I_s=16.5$ W/m$^2$, $\gamma=37$ MHz, $\omega_{0x}= 2\pi\times 10$ Hz, $\omega_{0y}=\omega_{0z}=2\pi\times 70$ Hz and $a_s=100a_0=5.29\times 10^{-11}$ m.}
	\label{f402}
\end{figure}

\begin{figure}
	\centering
	\includegraphics[width=8cm, height=6.5cm,angle=0] {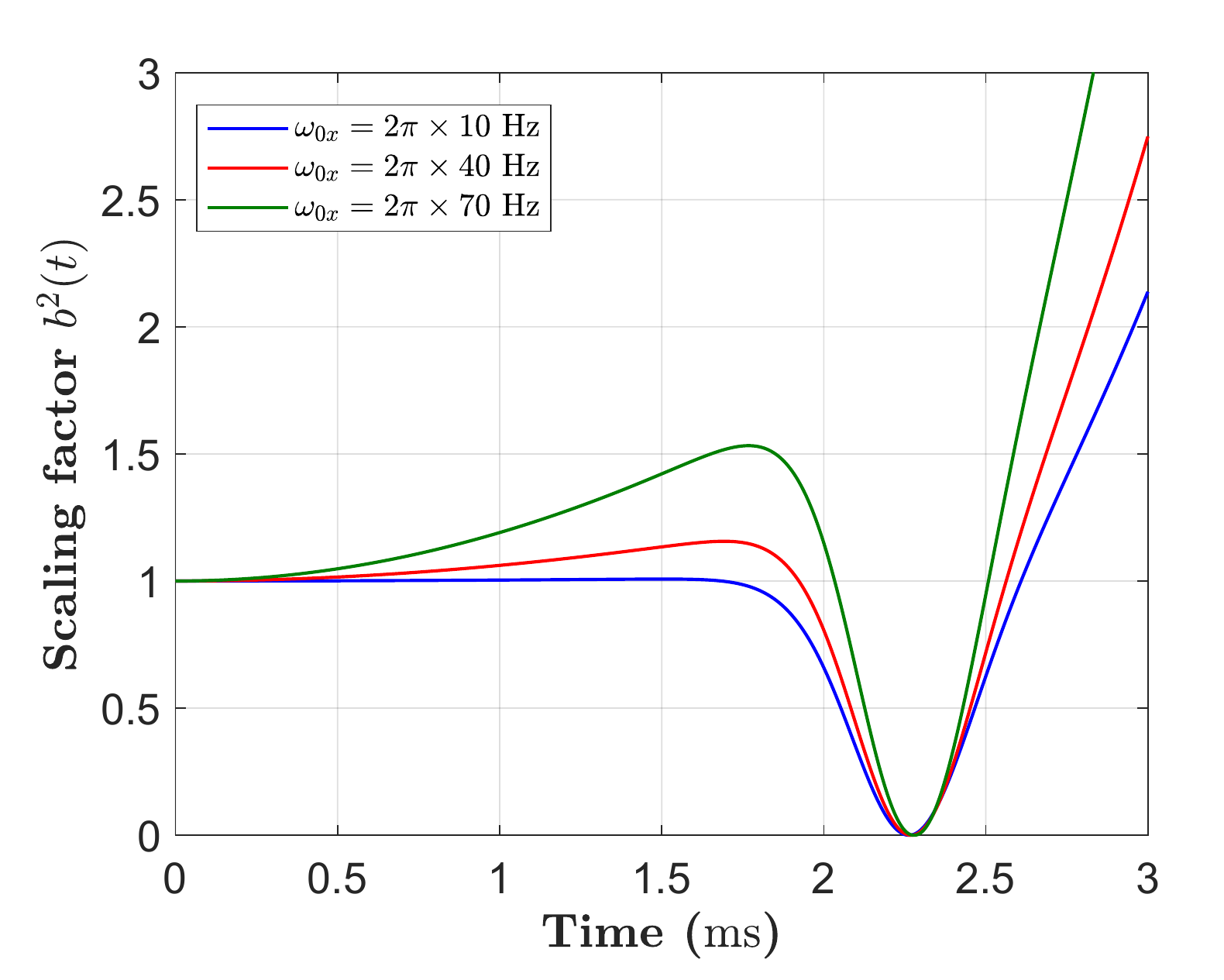}
	\caption{Numerical solution for the scaling factor, $b_x^2(t)$, as a function of time for three different axial trapping frequencies, $\omega_{0x}=2\pi\times 10$ Hz (blue curve), $\omega_{0x}=2\pi\times 40$ Hz (red curve) and $\omega_{0x}=2\pi\times 70$ Hz (green curve). The parameters are: $z_0=500~\mu$m, $\sigma_z=100~\mu$m, $\lambda=312~\mu$m, $\Delta=200$ GHz, $I_s=16.5$ W/m$^2$, $\gamma=37$ MHz and $\omega_{0y}=\omega_{0z}=2\pi\times 70$ Hz.}
	\label{f403}
\end{figure}

\begin{figure}
	\centering
	\includegraphics[width=9.5cm, height=10.5cm,angle=0] {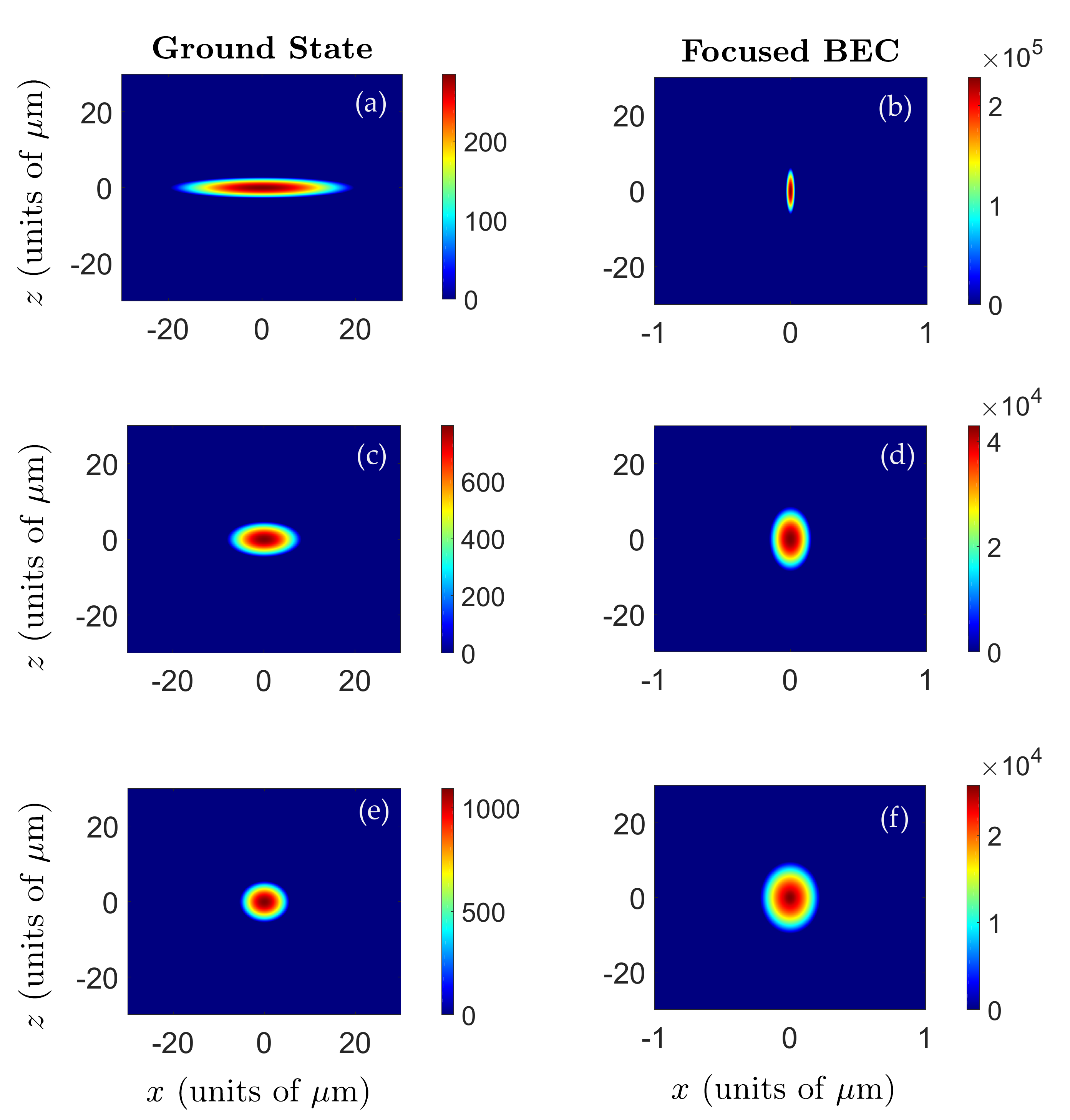}
	\caption{Left column: Top view of the TF ground state density profile in the ($x-z$) plane for (a): $\omega_{0x}=2\pi\times 10$ Hz, (c): $\omega_{0x}=2\pi\times 40$ Hz and (e): $\omega_{0x}=2\pi\times 70$ Hz. Right column: The resultant focused BEC density profile estimated by the scaling solution method for (b): $\omega_{0x}=2\pi\times 10$ Hz, (d): $\omega_{0x}=2\pi\times 40$ Hz and (f): $\omega_{0x}=2\pi\times 70$ Hz. The color maps in all graphs illustrate the values of density profile in atoms/$\mu\text{m}^2$. The parameters are: $N_0=10^5$, $z_0=500~\mu$m, $\lambda=312~\mu$m, $\sigma_z=100~\mu$m, $\Delta=200$ GHz, $I_s=16.5$ W/m$^2$, $\gamma=37$ MHz, $\omega_{0y}=\omega_{0z}=2\pi\times 70$ Hz and $a_s=100a_0=5.29\times 10^{-11}$ m for $^{87}$Rb.}
	\label{f404}
\end{figure}

	\section{Results}
	\label{sec:S4}
	
We now conduct a number of simulations using the scaling solution method to study the condensate evolution under a focusing harmonic potential. We suppose a cylindrical $^{87}$Rb BEC with $N_0=10^5$ confined in a harmonic trap including $\omega_{0x}=2\pi\times 10$ Hz, $\omega_{0y}=\omega_{0z}=2\pi\times 70$ Hz. The ground state of BEC inside the trap is obtained assuming that the condensate is highly repulsive having the scattering length of $a_s=100a_0$ ($a_0$ is the Bohr Radius, $a_0=5.29\times 10^{-11}$ m). The wavelength of the focusing potential, apart by $z_0=500~\mu$m from the BEC, is taken as $\lambda= 312~\mu$m (400 times greater than the actual $D_2$ line transition of $^{87}$Rb, $\lambda_{\text{D}_2}=780.027$nm). 
By adjusting the potential power appropriately, we aim that the BEC reaches its maximum focused state (optimal focus) at the center of the focusing potential \cite{RS_7}.

Fig.~\ref{f401} indicates a numerical solution of Eqs. (\ref{432}) and (\ref{433}) for the scaling factors, $b_x(t)$ and $b_y(t)=b_z(t)$ in the time interval $0\leqslant t\leqslant 3$ ms when the BEC is exposed to a time-dependent harmonic potential whose frequency follows Eq. (\ref{456}). Here, three different momentum kicks, $p=32\hbar k,\ 64\hbar k$ and $128\hbar k$, are initially applied to the condensate. The required power to bring the BEC to its optimum focus at the center of the potential ($z=0$) is determined by treating the atoms as classical particle trajectories \cite{RS_5, RS_7,RS_8}
\begin{equation}
	P_0=(5.37)\frac{\pi}{4}\frac{E_0}{\hbar\Delta}\frac{\gamma^2+4\Delta^2}{\gamma^2}\frac{I_s}{k^2},
	\label{e29}
\end{equation}
where $E_0$ represents the condensate's kinetic energy at $z=0$. Thus, the corresponding powers are estimated as $P_{32\hbar k}=19.544$ mW, $P_{64\hbar k}=65.532$ mW and $P_{128\hbar k}=249.481$ mW for the final velocities of $[v(z=0)]_{32\hbar k}=21.315$ cm/s, $[v(z=0)]_{64\hbar k}=39.031$ cm/s and $[v(z=0)]_{128\hbar k}=76.156$ cm/s respectively. As a reference, we set $\sigma_z=100~\mu$m, $\Delta=200$ GHz, $I_s=16.5$ W/m$^2$, $\gamma=37$ MHz (based on the data from the $^{87}$Rb D$_2$ line). Since the focusing potential is applied along the $x$ direction, it causes $b_x^2(t)$ to become very close to zero at a certain time depending on the BEC momentum kick and the corresponding potential power. As shown in Fig.~\ref{f401}, increasing the initial momentum kick yields the minimum point for $b^2_x$ curves at earlier times such that this occurs at $[t(z=0)]_{32\hbar k}=2.262$ ms, $[t(z=0)]_{64\hbar k}=1.266$ ms and $[t(z=0)]_{128\hbar k}=0.654$ ms for $p=32 \hbar k$, $p=64 \hbar k$ and $p=128 \hbar k$ respectively, shown by the solid blue, red and green lines. The dashed lines illustrate the evolution of the scaling factors, $b_y^2(t)$ and $b_z^2(t)$, along the radial axes in which no focusing dipole force is applied. Therefore, the three dashed curves are of ascending trend over the condensate propagation process due to the presence of repulsive \textit{s}-wave interactions. In addition, this trend undergoes a considerable boost for all the radial curves right at their $t(z>0)$ where the condensate leaves the focus point inside the potential and begins to expand significantly to release saved potential energy. Utilizing the values for the axial and radial scaling factors at the focus point ($z=0$), one is able to obtain the 3D density profile for the focused BEC at $z=0$ using Eq. (\ref{451}).

The results for the values of condensate Full Width at Half Maximum (FWHM) and peak density for different momentum kicks, from $p=0\hbar k$ to $p=128\hbar k$, are represented in Fig. \ref{f402}. We have considered two various potential radii, $\sigma_z=100$, $200~\mu$m to investigate the focused profile. Every momentum kick requires a certain optimum potential power to focus the BEC at $z=0$. For each potential radius, an increase in the magnitude of momentum kick [and consequently power, see Eq. (\ref{e29})] results in a lower profile FWHM improving the resolution of the created structures. A rise in potential power also enhances the exerted dipole force to the BEC bringing higher peak densities. As a case in point, imparting $p=128\hbar k$ to the BEC, respectively, leads to a resolution and peak density of $7.29$ nm and $1.682\times 10^6$ atoms/$\mu\text{m}^2$ in the plane of $x-z$ (the profile has been integrated over the $y$ axis), for $\sigma_z=100~\mu$m. Moreover, since increasing the potential size, $\sigma_z$, causes a smaller laser intensity according to $I_0=8P_0/\pi\sigma_z^2$ \cite{2_7, RS_7}, it is expected that this negatively affects the profile resolutions and lowers the peak densities of the focused condensate (see the bottom graph in Fig. \ref{f402}).

It is also worth investigating the influence of different BEC geometries on the deposited profile. Here, we consider three various cases in which the condensate is initially trapped by using three different axial trapping frequencies, $\omega_{0x}=2\pi\times 10$, $2\pi\times 40 $ and $2\pi\times 70$ Hz forming a highly cylindrical, moderate cylindrical and spherical BECs respectively, assuming that $\omega_{0r}=2\pi\times 70$ Hz is the same for all the three cases. The numerical results of solving $b_x^2(t)$, are illustrated in Fig. \ref{f403}. For $\omega_{0x}= 2\pi\times 70$ Hz, the variation slope of scaling curve is more significant than that of $\omega_{0x}= 2\pi\times 40$ Hz and $\omega_{0x}= 2\pi\times 10$ Hz, both before ($t<2.262$ ms) and after ($t>2.262$ ms) the focal spot. The reason for this is due to the greater amount of the potential energy saved in the confined BEC. This potential energy is converted to the kinetic energy when the BEC is released from the trap as well as when it leaves the focal spot. Moreover, while the scaling factor, $b_x^2(t)$, has the largest values for $\omega_{0x}= 2\pi\times 70$ Hz, it takes the lowest values for $\omega_{0x}= 2\pi\times 10$ Hz, in $0\leqslant t\leqslant 3$ ms so that $(b_x^2)_{2\pi\times 70\text{Hz}}>(b_x^2)_{2\pi\times 40 \text{Hz}}>(b_x^2)_{2\pi\times 10 \text{Hz}}$.

The results at $t=0$ (BEC ground state) and $t=2.262$ ms (focused state) are applied to the BEC density profile [Eq. (\ref{451})], and they are depicted in Figs. \ref{f404} (a-f). While the left column graphs, from top to bottom [Figs. \ref{f404} (a), (c) and (e)], indicate the BEC ground state for $\omega_{0x}=2\pi\times\ 10,\ 40,\ 70$ Hz respectively, the right column ones [Figs. \ref{f404} (b), (d) and (f)] show the corresponding focused BEC at $t(z=0)=2.262$ ms. It is clear that increasing the axial trapping frequency whilst keeping the radial one constant, results in a wider (less resolution) and shorter (less peak density) focused condensate as this brings smaller $b_x^2$. Hence, it is expected that focusing a cigar-shaped BEC leads to a higher resolution and peak density compared to a spherical BEC given that a cylindrical BEC is elongated along the direction where the dipole force is applied (the $x$ axis in this case).

\begin{figure}[t!]
	\centering
	\includegraphics[width=9cm, height=12.5cm,angle=0] {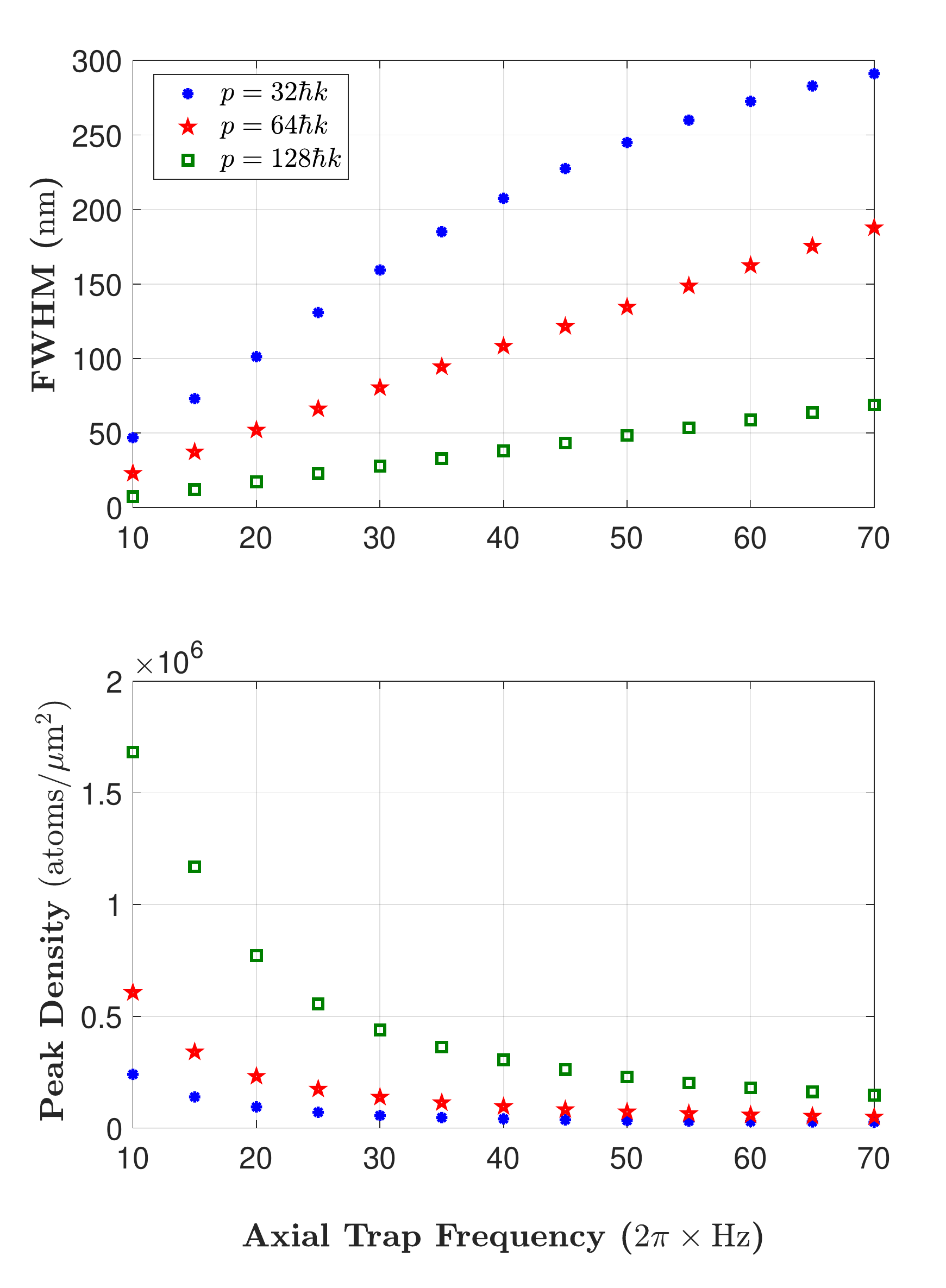}
	\caption{Results for the focused $^{87}$Rb condensate at $z=0$ derived from the scaling solution model for different values of axial trapping frequency. The simulations are conducted using three various momentum kicks, $p=32\hbar k$ (blue dots), $64\hbar k$ (red triangles) and $128\hbar k$ (green squares). The upper graph indicates the values of FWHM while the lower one shows the corresponding peak densities. Parameters used are: $N_0=10^5$, $z_0=500~\mu$m, $\sigma_z=100~\mu$m, $\lambda=312~\mu$m, $\Delta=200$ GHz, $I_s=16.5$ W/m$^2$, $\gamma=37$ MHz, $\omega_{0y}=\omega_{0z}=2\pi\times 70$ Hz and $a_s=100a_0=5.29\times 10^{-11}$ m.}
	\label{f405}
\end{figure}

Following the consideration of various BEC geometries, we have considered the profile characteristic factors (FWHM and peak density) in the range of $2\pi\times 10\leqslant \omega_{0x}\leqslant 2\pi\times 70$ Hz for $p=32$, $64$ and $128\hbar k$, shown in Figs. \ref{f405}. As observed, narrower and higher structures are achieved when utilizing lower axial trapping frequencies for any magnitude of $p$. Furthermore, the influence of employing larger momentum kicks on the improvement of focused structures is displayed, which offers the best results for $p=128\hbar k$.

\section{Numerical GPE Simulations}

In order to validate the accuracy of the analytical scaling solution approach in estimating the dynamics of focusing BECs, we performed the exact GPE numerical solutions. To this end, we numerically solved Eq.(\ref{429}) using an embedded Runge-Kutta (ERK) method in conjunction with adaptive Fourier split-step size \cite{43, RS_7}. In this process, the third and fourth orders [ERK4(3)] were employed, evaluating a local error for an adaptive step-size control in each iteration of a simulation. The characteristic factors of the focused BEC profile through a harmonic focusing potential were examined by the numerical GPE simulations. The results indicate that the scaling solution method, in most cases, offers a reliable precision in predicting the deposited structure linewidths and peak densities. 

\begin{figure}[ht]
	\centering
	\includegraphics[width=9cm, height=12.5cm,angle=0] {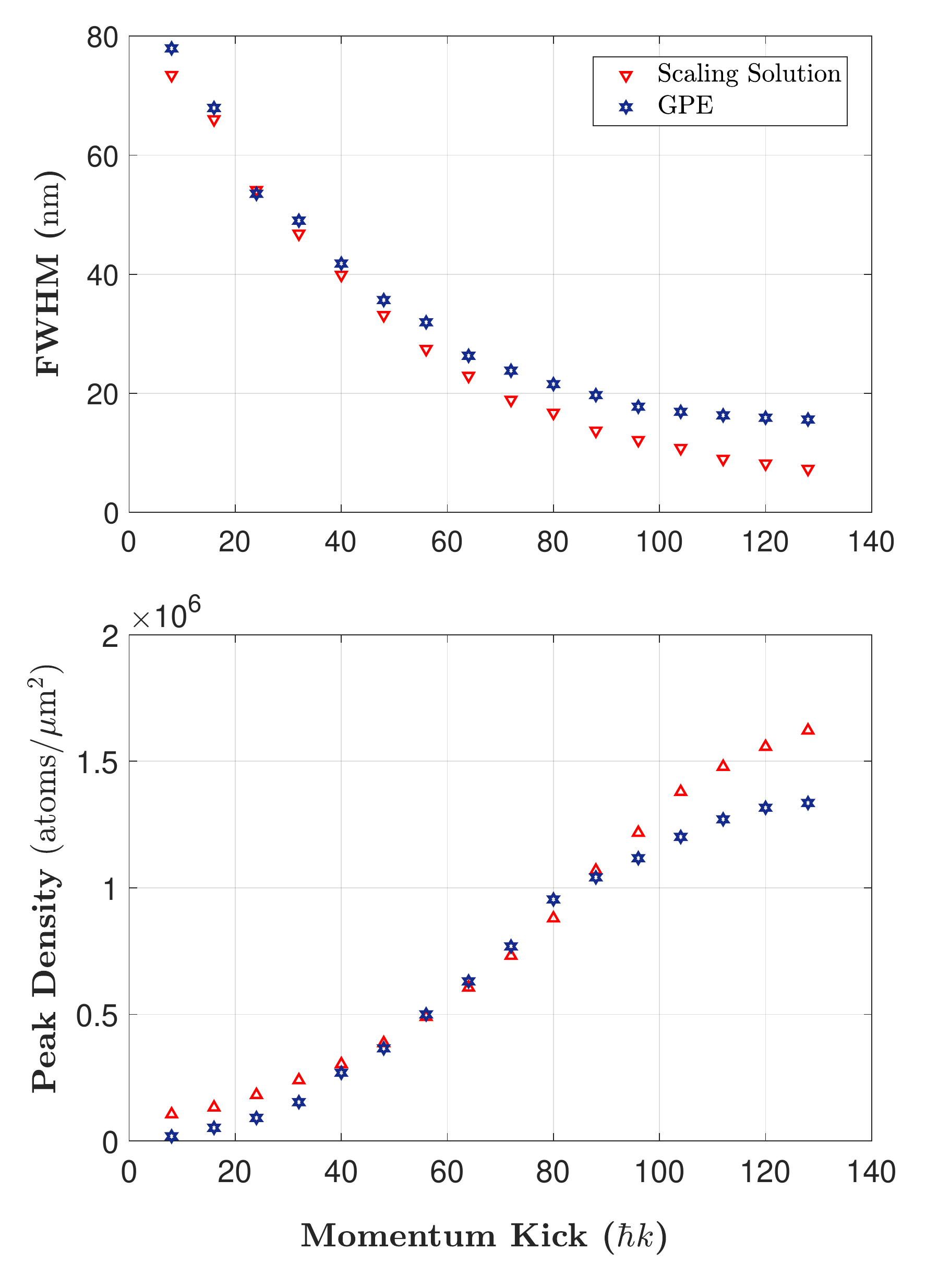}
	\caption{Results for the focused $^{87}$Rb condensate at $z=0$ achieved by the scaling solution (indicated by the red triangles) and exact numerical GPE simulation (indicated by the blue hexagrams) models. The upper graph illustrates the values of FWHM as a function of momentum kick whereas the lower graph shows the corresponding peak densities. Parameters used are: $N_0=10^5$, $z_0=500~\mu$m, $\sigma_z=100~\mu$m, $\lambda=312~\mu$m, $\Delta=200$ GHz, $I_s=16.5$ W/m$^2$, $\gamma=37$ MHz, $\omega_{0y}=\omega_{0z}=2\pi\times 70$ Hz and $a_s=100a_0=5.29\times 10^{-11}$ m.}
	\label{f2}
\end{figure}

Fig. \ref{f2} displays the profile resolution and peak density as a function of imparted momentum kick acquired by the scaling solution and numerical GPE approaches for the same parameters as in Section~\ref{sec:S4}. Concentrating on the profile width, we notice that the scaling solution (red triangle curve) delivers higher accuracy in the results for the lower momentum kicks (i.e. $p< 48 \hbar k$). However, according to the GPE simulations, the resolution tends to a steady state for $p> 100 \hbar k$, which causes a gap between the blue and red curves. As an illustration, we achieved the linewidth of $(\Delta x)_{\text{res}}= 15.6$ and $7.3$ nm for $p= 128\hbar k$ out of the GPE and scaling solution model respectively.

In regard to the profile peak density, we found that the best agreement occurs when setting $40\leqslant p\leqslant 88\hbar k$. Nonetheless, for extremely high momentum kicks, the scaling solution methodology loses its capability in providing the precise results such that a considerable offset is observed between the blue and red curves at $p=128\hbar k$. This is mainly due to the excitations in the focusing event \cite{RS_7} that become significant when applying larger momentum kicks to the BEC, which can be predicted well through the GPE simulations.

\section{Conclusions}

In this paper, we investigated the dynamics of focusing of 87Rb BECs travelling through a harmonic-shaped focusing optical potential from an analytical point of view. Utilizing the scaling solution technique, we analyzed the evolution of 87Rb	condensates in the expanding and focusing regimes when considering the inter-atomic two-body interactions. We used a highly repulsive BEC (i.e. $a_s =100a_0$) satisfying the requirement for a Thomas-Fermi profile. It was concluded that the initial geometry of a BEC can play an essential role in the focused scheme so that better resolutions and peak densities can be achieved by cylindrical BECs rather than spherical ones. We showed that using higher potential powers and initial momentum kicks as well as exploiting smaller potential radius sizes, can significantly improve the resolution and peak density of deposited structures. Finally, a direct comparison between the scaling solution and numerical GPE simulations was conducted. We found a good agreement between the two approaches, specially, for low and mid-range longitudinal velocities. Nevertheless, since the proposed analytical method does not consider the dimensionality disruptions such as excitations in high momentum kick regimes, the results may not be as reliable as in the slow regimes.\\ \\

\section*{ACKNOWLEDGMENTS}

R.R. is supported by the Australian National University International Research Scholarship award. The authors would like to thank Timothy Senden and Hans A. Bachor for useful discussions and feedback.

	\bibliography{RS_1,RS_2,RS_3,R2_17,Ref_3,Ref_54,RI_41,RI_42,RI_43,RI_44,RI_45,RI_46,RI_47,RI_48,RI_49,RI_50,RI_51,RI_40,R3_2,R3_3,RI_75,RI_6,RI_74,RS_4,RS_5,RS_6,RS_7,RS_8,R4_1,R4_3,R3_5,R2_7,R2_2,R2_6,Ref_38,Ref_39,Ref_43,Ref_37}
\end{document}